\documentclass[aps,prl,twocolumn,floatfix,english,showpacs,10pt,superscriptaddress]{revtex4-2}%
\usepackage{graphicx}
\usepackage{amsmath}
\usepackage{physics}
\usepackage{amssymb}
\usepackage{mathrsfs}
\usepackage{colordvi}
\usepackage{verbatim}
\usepackage{xcolor}
\usepackage{mathrsfs}
\usepackage{epsfig}
\usepackage{lipsum}
\usepackage{amsfonts}
\usepackage[unicode=true, breaklinks=false, pdfborder={0 0 1}, backref=false,
colorlinks=true, linkcolor=blue, urlcolor=blue, citecolor=blue]{hyperref}%
\newcommand{\ii}{i}
\newcommand{\ee}{e}

\providecommand{\U}[1]{\protect\rule{.1in}{.1in}}

\setcitestyle{numbers,square}
\begin{document}
\title{Abnormal Frequency Response Determined by Saddle Points in Non-Hermitian Crystal Systems}
\author{Kunling Zhou}
\thanks{These authors contribute equally to this work.}
%\altaffiliation{}

\affiliation{Hunan Provincial Key Laboratory of Flexible Electronic Materials Genome Engineering,
School of Physics and Electronic Sciences, Changsha University of Science and Technology, Changsha 410114, P. R. China}
\affiliation{School of Physics, Huazhong University of Science and Technology, Wuhan 430074, P. R. China}

\author{Jun Zhao}
\thanks{These authors contribute equally to this work.}
\affiliation{School of Physics, Huazhong University of Science and Technology, Wuhan 430074, P. R. China}

\author{Bowen Zeng}
\email[]{zengbowen@csust.edu.cn}
\affiliation{Hunan Provincial Key Laboratory of Flexible Electronic Materials Genome Engineering,
School of Physics and Electronic Sciences, Changsha University of Science and Technology, Changsha 410114, P. R. China}
\affiliation{School of Physics, Huazhong University of Science and Technology, Wuhan 430074, P. R. China}

\author{Yong Hu}
\email[]{huyong@hust.edu.cn}
\affiliation{School of Physics, Huazhong University of Science and Technology, Wuhan 430074, P. R. China}

\makeatletter
\newcommand{\rmnum}[1]{\romannumeral #1}
\newcommand{\Rmnum}[1]{\expandafter\@slowromancap\romannumeral #1@}
\makeatother

\begin{abstract}
In non-Hermitian crystal systems under open boundary conditions (OBCs), it is generally believed that the OBC modes with frequencies containing positive imaginary parts, when excited by external driving, will experience exponential growth in population, thereby leading to instability. However, our work challenges this conventional understanding. In such a system, we find an anomalous response that grows exponentially with the frequencies aligned with those of saddle points. The frequencies of these saddle points on the complex plane are below the maximum
imaginary part of the OBC spectrum, but they can lie within or beyond the OBC spectrum. We derive general
formulas of excitation–response relationships and find that this anomalous response can occur because the excitation of OBC modes eventually evolves toward these saddle points at long times. Only when the frequencies of all these saddle points are below the real axis do the non-Hermitian crystal systems remain stable under periodic excitation. Thus our results also provide insights on the stability criterion of non-Hermitian crystal systems.

\end{abstract}
\maketitle

\textit{Introduction.}---The study of response to excitation is a central issue in many physical models, ranging from forced vibration in classic mechanics~\cite{landau1975course} and  transient response in circuits \cite{signal1996} to the dynamic evolution of wavefunction in quantum physics \cite{Griffiths_Schroeter_2018}. For a linear control system subjected to periodic excitation, it is generally believed that the response frequency is intimately connected  to the excitation frequency, the system's natural frequencies, or a combination of both~\cite{landau1975course,signal1996}. This understanding also uncovers the stability criteria~\cite{hurwitz1895ueber,1102280}. For a linear control system to be stable, the natural frequencies may have imaginary parts but these imaginary parts must not be greater than zero; otherwise, the natural response would grow exponentially over time, thereby undermining the system's stability.

\begin{figure}
    \centering
    \includegraphics[width = 8.6 cm]{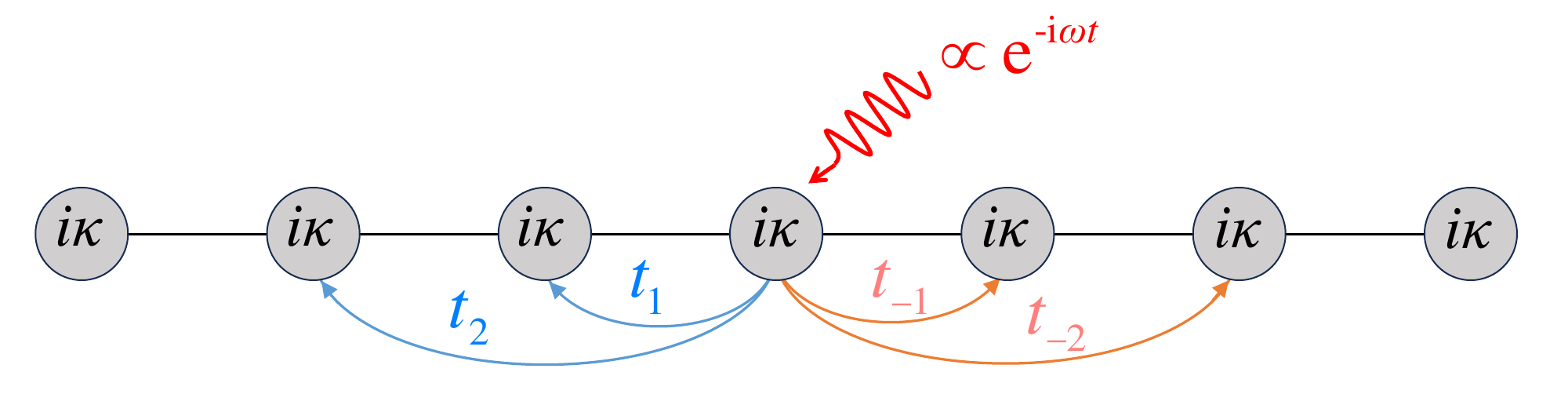}
    \caption{Schematic diagram of a non-Hermitian crystal system  undergoing periodic excitation $e^{-i \omega t}$. Here, the non-Hermiticity is characterized by on-site dissipation $i \kappa$ and non-conjugated hopping $t_{n}\neq t_{-n}^{*} $, where $t_{n}$ represents  the hopping strength from site $i+n$ to site $i$ (for any $i$).  }
    \label{fig-model}
\end{figure}

In recent years, with the rapid development of non-Hermitian physics~\cite{doi:10.1080/00018732.2021.1876991,Gongzp2018topologicalphases,PhysRevLett.121.086803,yokomizo2019non,Fangchen2020Correspondence,Fangchen2020Correspondence,yang2020non,2020arXiv201203333Y,Brody_2014,RevModPhys.93.015005,helbig2020generalized,yu2024non,PhysRevLett.124.056802}, the issue of excitation–response relationships in non-Hermitian crystal systems, as shown in Fig.~\ref{fig-model}, has attracted extensive attention~\cite{li2021quantized,Wanjura2020topological,Wanjura2021disorder,Wangzhong2021Simpleformulas,PhysRevX.8.041031,mcdonald2020exponentially,hashemi2022linear,PhysRevLett.122.143901,2024arXiv240510176V,Tian:23,hashemi2022linear,zhang2021acoustic}. The Green's function is a crucial tool for studying this issue and the
explicit formulas of its non-Hermitian version~\cite{zirnstein2021bulk,zirnstein2021exponentially,Wangzhong2021Simpleformulas,xue2022non,BCindependence2021,PhysRevResearch.5.043073,PhysRevB.107.115412,PhysRevLett.124.056802} recently proposed by Xue \textit{et al.}~\cite{Wangzhong2021Simpleformulas} revealed a potential application of non-Hermitian systems as directional amplifiers. This non-Hermitian version primarily focused on the response from the external driving, also known as forced response, which is appropriate when large on-site dissipation is applied to suppress the contribution from the natural response~\cite{Wangzhong2021Simpleformulas}. However, without the constraint of Hermiticity, the natural frequencies of non-Hermitian crystal systems under OBC such as OBC spectrum can possess positive imaginary parts even in lossy lattices~\cite{zhang2021acoustic}. In such a case, a recent quantum walk experiment~\cite{PhysRevLett.126.230402} indicated that the dynamic evolution of probabilities of the photon is dominated by the frequency with the largest imaginary part in the OBC spectrum. However, another study~\cite{PhysRevResearch.1.023013} by Longhi pointed out that the bulk dynamic evolution of initial states is limited by an upper bound determined by spectrum under periodic boundary condition. There is still some controversy regarding the excitation–response relationships of non-Hermitian systems, and to the best of our knowledge, a unified description has yet to be established. 

In this Letter, for a non-Hermitian crystal system subjected to periodic excitation, we observe an anomalous exponentially growing response, with its frequency aligned with the frequency of a saddle point beyond the range of OBC spectrum. By utilizing the Laplace transform in the dynamic equation, we derive general formulas of excitation–response relationships, successfully explaining this phenomenon and clarifying the aforementioned controversy. Our formulas demonstrate that the complete response can be divided into forced response by external excitation and natural response by natural frequencies, but the long-time behavior of complete response is determined by their competition. Intriguingly, the frequency components of this natural response are initially composed of the OBC spectrum but eventually evolve to the frequencies of saddle points. These saddle points' frequencies can lie within or beyond the OBC spectrum. From this, we infer that for a non-Hermitian crystal system to be stable undergoing excitation, all of these saddle points must be located below the real axis.

\begin{figure}
    \centering
    \includegraphics[width = 8.6 cm]{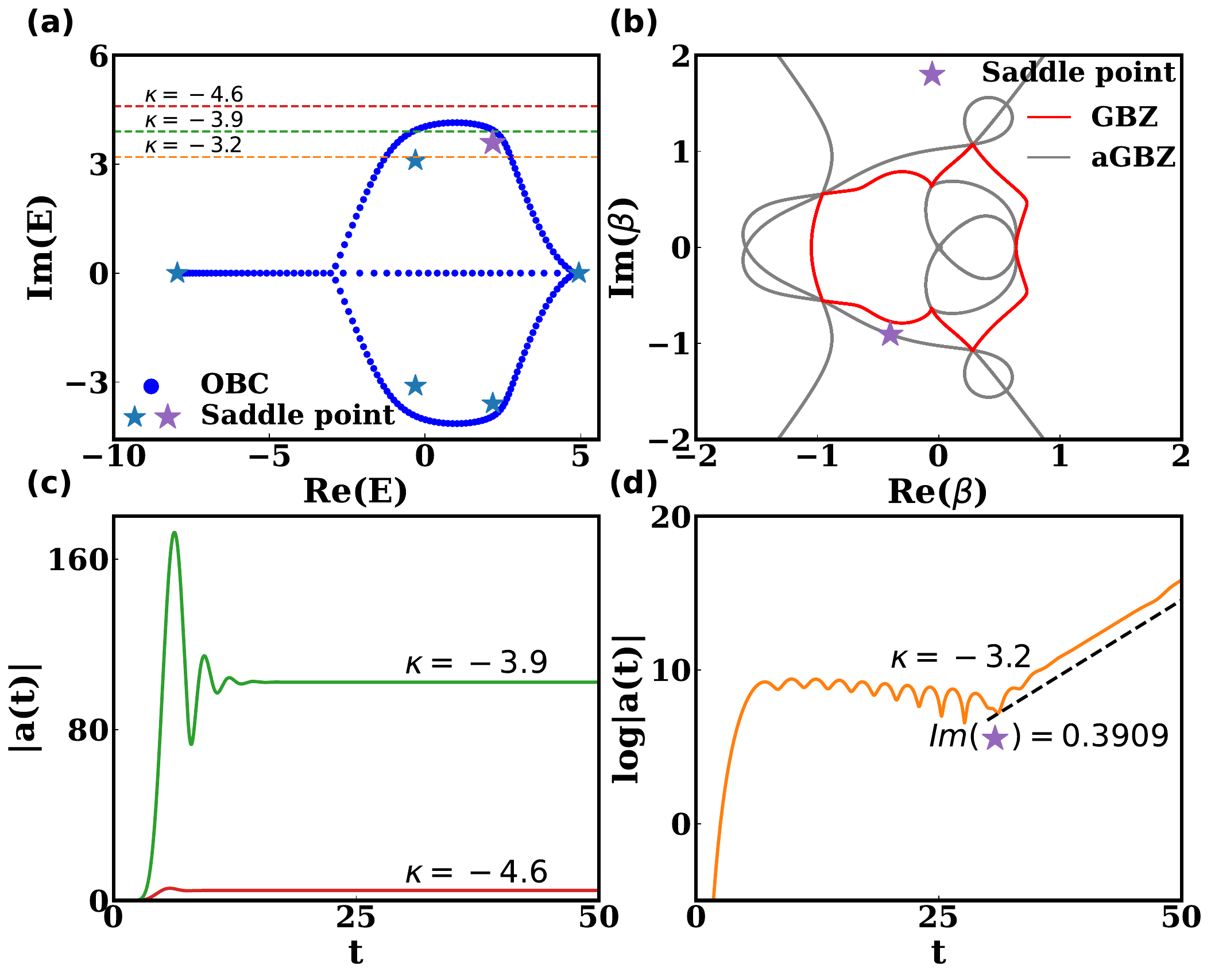}
    \caption{ Concrete model with parameters $\kappa = 0$, $t_{3}=3, t_{2}=t_{1}=t_{-1}=2, t_{-2}=-2$, and $t_{-3}=1$  is illustrated, with the numerical calculations of OBC spectrum for $N=1600$ sites shown in (a) and GBZ and aGBZ presented in (b). In these diagrams, the saddle points are represented by pentagram, while the uppermost saddle point (USP) with the largest imaginary part is highlighted by purple pentagram in (a) and the corresponding wavevector is shown in (b). Under excitation at site 800, the evolution of amplitude  $\left|a(t)\right|$ at site 750 is depicted with different on-site dissipation: (c) $\kappa = -4.6$, $\kappa=-3.9$ and (d)  $\kappa = -3.2$. The amplitude in (d) exhibits exponential growth with frequency aligned with the frequency of the USP in (a).   }
    \label{fig-2}
\end{figure}

\textit{Abnormal frequency response.}---The schematic diagram of a non-Hermitian crystal system subjected to periodic excitation $e^{-i \omega t}$ with $\omega$ being the excitation frequency is shown in Fig.~\ref{fig-model}. We first attempt to exhibit counterintuitive results in Fig.~\ref{fig-2} using a concrete non-Hermitian model, which takes a non-Bloch Hamiltonian
\begin{equation}\label{eq-model}
    h(\beta)=3 \beta^3+2 \beta^2+2 \beta+ i\kappa + 2 \beta^{-1}-2 \beta^{-2}+\beta^{-3},
\end{equation}
where $\kappa$ is the on-site uniform dissipation. Here, $\beta$ is the allowed wavevector under OBC and the corresponding frequencies are the OBC spectrum $E_{OBC}$. The allowed wavevectors and OBC spectrum of the non-Hermitian systems are dramatically different from those under periodic boundary condition due to their striking sensitivity to the boundary~\cite{PhysRevLett.121.026808,PhysRevLett.116.133903,PhysRevLett.127.116801,PhysRevB.99.201103}. When $\kappa=0$, Fig.~\ref{fig-2}(a) shows the complex OBC spectrum, which is symmetric about the real axis. The modes below the real axis have negative imaginary parts that denote a finite lifetime \cite{PhysRevX.7.031024,PhysRevLett.122.203605}. In contrast, the modes with positive imaginary parts typically imply their population grows exponentially over time~\cite{PhysRevA.99.063834}. 

 The OBC spectrum $E_{OBC}$ becomes continuous for infinite-size non-Hermitian crystal systems while to form continuum spectrum, the allowed wavevectors, i.e., the roots of characteristic function $  E_{OBC} - h(\beta) =-3 \prod_i \left(\beta - \beta_{i} \right)/\beta^3 =0$ must fulfill the condition $\left|\beta_3 \right|=\left|\beta_4 \right|$, where $\beta_{i}$ and $\beta_{3, 4}$ label the $i$-th root and $3, 4$-th root sorted by their moduli, respectively~\cite{yokomizo2019non,PhysRevLett.121.086803}. The collection of these allowed wavevectors on the complex plane is the so-called generalized Brillouin zone (GBZ), as shown in Fig.~\ref{fig-2}(b). A standard numerical method to calculate two roots with equal modulus typically yields the auxiliary GBZ (aGBZ) $\left|\beta_i \right|=\left|\beta_{j} \right|$ (for any $i,j$) that comprises the GBZ~\cite{Fangchen2020Correspondence,yang2020non}, as shown in Fig.~\ref{fig-2}(b).

Now we turn to the response as the periodic excitation is applied. The dynamic evolution for model described by Hamiltonian Eq.~\eqref{eq-model} is governed by~\cite{RevModPhys.82.1155,Wanjura2020topological,PRXQuantum.5.020202} 
\begin{equation}
    i \frac{d \mathbf{a(t)}}{d t} = H\mathbf{a(t)} + \mathbf{b}_{in} e^{- i\omega t}
     \label{eq-eom}
\end{equation}
where $H$ is the matrix form of the non-Hermitian Hamiltonian. Here, $\mathbf{a}(t) = (a_1(t), a_2(t), \cdots, a_N(t) )^T$ is the amplitude vector of the lattice and $\mathbf{b}_{in}$ is the input amplitude. Note that each $a_i(t)$ is just a complex number, which can be considered as the expectation value of the corresponding operator in the dynamic equation. We only consider the response with zero initial condition $\mathbf{a}(0)=0$ in the following analysis. Under excitation at site 800, the response amplitudes $\left|a(t)\right|$ at site 750 with different dissipation are shown in Figs.~\ref{fig-2}(c) and~\ref{fig-2}(d). Only when $\kappa = -3.2$ does the amplitude grow exponentially over time. A similar trend in the dynamic evolution of amplitude can be found at other sites as shown in Fig. S1 in the Supplemental Material~\cite{SupplementalMaterials}. With large on-site uniform dissipation $\kappa=-4.6$, which ensures all the OBC spectrum lies below the real axis [Fig.~\ref{fig-2}(a)], Fig.~\ref{fig-2}(c) shows that the evolution of $\left|a(t)\right|$ remains unchanged after a small initial increase. From Eq.~\eqref{eq-eom}, the excitation $\mathbf{b}_{in}e^{- i\omega t}$ can be decomposed by a superposition of eigenvectors, thus leading to the activation of a series of the OBC spectrum. However, this natural response decays with time due to their negative imaginary energy while the forced response survives under long-time evolution. Altering the dissipation to $\kappa = -3.9$ renders the maximum imaginary value of the OBC spectrum [denoted by $\operatorname{Im}\left(E_m\right)$] becoming greater than zero. This change leads to a rapid amplification of response amplitude, but which eventually stabilizes. These results suggest that the OBC modes with positive imaginary parts are indeed excited for a short period but then disappear. 

However, when $\kappa = -3.2$, Fig.~\ref{fig-2}(d) shows an exponential growth of amplitude with specified frequency after a short-term growth and oscillation. This growth rate is precisely equal to the imaginary part of
the uppermost saddle point [USP; represented by purple pentagram in Fig.~\ref{fig-2}(a)], which has maximum imaginary part among saddle points below the $\operatorname{Im}\left(E_m\right)$. Mathematically, the saddle points satisfy $\partial h(\beta) /\partial \beta |_{\beta_s}=0$, which also corresponds to the occurrence of multiple roots~\cite{PhysRevLett.132.050402}. Therefore, they are always located on the aGBZ, but they may not be found on the GBZ~\cite{PhysRevLett.132.050402}. The frequency of USP neither lies within the OBC spectrum nor aligns with the external driving frequency. Therefore, this exponential growth amplitude is referred to as abnormal frequency response. 

\textit{General formulas of excitation–response relationships.}---To gain deeper insight into the abnormal frequency response mentioned above, we turn to deriving the excitation–response relationships for a general non-Hermitian Hamiltonian $h(\beta) = \sum_{n=-l}^{r} t_{n} \beta^{n}$. 
Considering the possibility of exponential growth in amplitude, we first introduce the Laplace transform in Eq.~\eqref{eq-eom}
\begin{equation} 
    \mathbf{a}(p)=\frac{1}{i p -H}\frac{1}{p+i \omega}\mathbf{b}_{in},
    \label{eq-ap}
\end{equation}
where $p$ is a complex variable that contains real $s>s_0$ to ensure convergence with $s_0$ being the abscissa of absolute convergence. 
The relationship between the response and excitation in the frequency domain is thus given by the frequency domain of the Green's function, multiplied by the Laplace transform of the initial conditions
\begin{align}
    G(p)=\frac{1}{i p -H}\frac{1}{p+i \omega},
\end{align}
with the matrix element $[G(p)]_{jk}$ denoting the response at site $j$ to the excitation at site $k$. For an infinite-size system, such matrix can be given by inverting Toeplitz matrices as proposed by Xue \textit{et al.}~\cite{bottcher2012introduction,Wangzhong2021Simpleformulas,10.5555/1121654}
\begin{equation}
\label{Eqn GK}
        [G(p)]_{jk} = \frac{1}{2\pi i} \oint_{\beta=GBZ} \frac{\beta^{j-k-1}}{i p - h(\beta)} \frac{1}{p+i \omega} \mathrm{d} \beta,
\end{equation}
where the integral loop is proven to the GBZ~\cite{Wangzhong2021Simpleformulas}. Then the time evolution of the system can be obtained through inverse Laplace transform
\begin{align}
    &[G(t)]_{jk}= \nonumber \\
    &\left(\frac{1}{2 \pi i}\right)^2   \int_{s-i \infty}^{s+i\infty} \oint_{\beta=GBZ} \frac{\beta^{j-k-1}}{i p - h(\beta)} \frac{1}{p+i\omega} e^{p t}  \mathrm{d} \beta \mathrm{d} p .
    \label{eq-inverseLT}
\end{align}
This integral can be evaluated by calculating the residue at pole $i p - h(\beta)=0$, which requires expressing the root as a function of $p$. For a simple Hatano-Nelson model~\cite{PhysRevB.58.8384}, the explicit analytical solution of dynamic evolution is derived, which is consistent with the numerical calculations as detailed in the Supplemental
Material~\cite{SupplementalMaterials}. However, when $h(\beta)$ involves long-range couplings beyond nearest neighbors, expressing the root as a function of $p$ can be challenging or impossible. To address this challenge, we change the order of integration, which simplifies the evaluation of this integral and makes the physical interpretation explicit, as discussed next.

Clearly, there are two kinds of poles for variable $p$ that contribute to the response. The contribution from pole $p+i\omega =0$ arises from the  external driving 
\begin{equation}\label{eq-Eterm}
    [G(t)]_{jk}(E)=\frac{1}{2 \pi i}\oint_{\beta=GBZ}\frac{\beta^{j-k-1}}{\omega - h(\beta)} e^{-i\omega t}\mathrm{d}\beta, 
\end{equation}
which corresponds to the forced response. By further investigating the root $w - h(\beta) =\prod_i -t_r \left(\beta - \beta_{i} (\omega)\right)/\beta^{l}=0$ with $\beta_{i} (\omega)$ being the $i$-th root sorted by their moduli, we have 
\begin{align}\label{eq-GBZ}
       &[G(t)]_{jk}(E) \nonumber \\
       =&\left\{
        \begin{aligned}
            &\sum_{n=1}^l \frac{-\beta_n(\omega)^{j-k+l-1}}{t_r\prod_{k\neq n}[\beta_n(\omega)-\beta_k(\omega)]}\ee^{ -i \omega t}, j-k+l-1\geqslant0, \\
            &\sum_{n=l+1}^{l+r} \frac{ \beta_n(\omega)^{j-k+l-1}}{t_r\prod_{k\neq n}[\beta_n(\omega)-\beta_k(\omega)]}\ee^{ -i \omega t}, j-k+l-1<0.
        \end{aligned} 
            \right.
\end{align} 
For the response sites far from the excitation site, the above equation can be simplified as $[G(t)]_{jk}(E) \sim \beta_l(\omega)^{j-k} $ when $ j\gg k $ and $[G(t)]_{jk}(E) \sim \beta_{l+1}(\omega)^{j-k} $ when $ j\ll k $. These results return to the explicit Green’s functions, as detailed in previous studies~\cite{Wangzhong2021Simpleformulas,PhysRevResearch.5.043073,xue2022non}. 
Now we turn to investigating the contribution from pole $ip-h(\beta)=0$
\begin{equation}\label{eq-Iterm}
    [G(t)]_{jk}(N)=\frac{1}{2 \pi i} \oint_{\beta=GBZ} -\frac{\beta^{j-k-1}}{\omega - h(\beta)} e^{-ih(\beta)t} \mathrm{d}\beta, 
\end{equation}
which represents the response arising from the natural frequencies, dubbed natural response. The natural response includes the contribution from all the frequencies of the system. This term may result in the exponential growth of amplitude since $h(\beta)$ can have positive imaginary parts. It should be noted that all the frequencies of the system are not equivalent to the OBC spectrum, though the integral zone is GBZ in  Eq.~\eqref{eq-Iterm}, since the frequencies may alter during time evolution. From the OBC spectrum, the upper limit of this integral can be determined,
\begin{align}
\abs{[G(t)]_{jk}(N)} \leqslant e^{\operatorname{Im}\left(E_m\right) t} \frac{1}{2 \pi } \oint_{\beta= \mathrm{GBZ}} \abs{\frac{\beta^{j-k-1}}{\omega - h(\beta)}}\abs{\mathrm{d}\beta},
\end{align}
since $\operatorname{Im}\left(h(\beta)\right)\leq \operatorname{Im}\left(E_m\right)$.

It should be noted the Laurent expansion of the exponential term in Eq.~\eqref{eq-Iterm} gives rise to infinite essential singularities, posing a challenge to residue theorem. At long times, the asymptotic expansion of Eq.~\eqref{eq-Iterm} can be obtained using the method of steepest descent from textbook~\cite{de1981asymptotic,ablowitz2003complex}. According to this method, the integral path can be designed to pass through all the saddle points and the asymptotic expansion is mainly contributed by the surrounding regions of these saddle points. Although deforming the integration path may across some pole  $h(\beta) = \omega$, this term introduces an oscillatory error with constant magnitude and can therefore be safely neglected. Consequently, the response by the natural frequencies becomes the sum of contributions from all these saddle points, yielding the asymptotic form of the time-domain excitation-response relationship as follows:

\begin{align}\label{eq-saddle}
    [G(t)]_{jk}(N) \sim \sum_{s=1}^S  \frac{({\beta_s})^{j-k-1}}{h(\beta_s)- \omega}e^{-\ii h(\beta_s)t}  \sqrt{\frac{i}{2\pi h^{\prime \prime}(\beta_s) t}}.
\end{align}
Here, $h^{\prime \prime}(\beta_s)$ is the second derivative~\cite{de1981asymptotic} at saddle points and $S$ represents the number of saddle points below $\operatorname{Im}\left(E_m\right)$. For simplicity, we assume that all the involved saddle points are second order and $h(\beta_s) - \omega \neq 0$ to avoid resonance. Equation~\eqref{eq-GBZ} and  Eq.~\eqref{eq-saddle} indicate the frequency components of natural response initially consist of the OBC spectrum and eventually evolve towards saddle points' frequencies at long times. It is expected that the USP will dominate the long-time dynamic behavior because it has the maximum positive imaginary part among all the saddle points. As pointed out earlier, such saddle points must lie on the aGBZ, but they may not necessarily lie on the GBZ.

It should be noted that the complete response [the sum of Eqs.~\eqref{eq-Eterm} and \eqref{eq-Iterm}] does not rely on the boundary conditions, provided that the response has not yet reached the boundary. This 
result aligns with the finding in Reference~\cite{BCindependence2021}. Therefore, the integral contour (GBZ) for the complete response can be replaced by the conventional Brillouin zone, though this transition in integral contour may alter both the integration of Eqs.~\eqref{eq-Eterm} and \eqref{eq-Iterm}, as explained in the Supplemental Material~\cite{SupplementalMaterials}. We also notice that $[G(\omega)]_{j k}$ obtained by Fourier transformation in Reference~\cite{Wangzhong2021Simpleformulas} and $[G(p)]_{j k}$ by Laplace transformation in our work depends on the boundary conditions, since such transformations involve integration over infinite time. In other words, infinite time suggests that the response should have reached the boundary and may continue to propagate, thereby allowing the boundary conditions to alter bulk element of $[G(p)]_{j k}$, as detailed in the Supplemental Material~\cite{SupplementalMaterials}. 

\begin{figure}
    \centering
    \includegraphics[width=8.6cm]{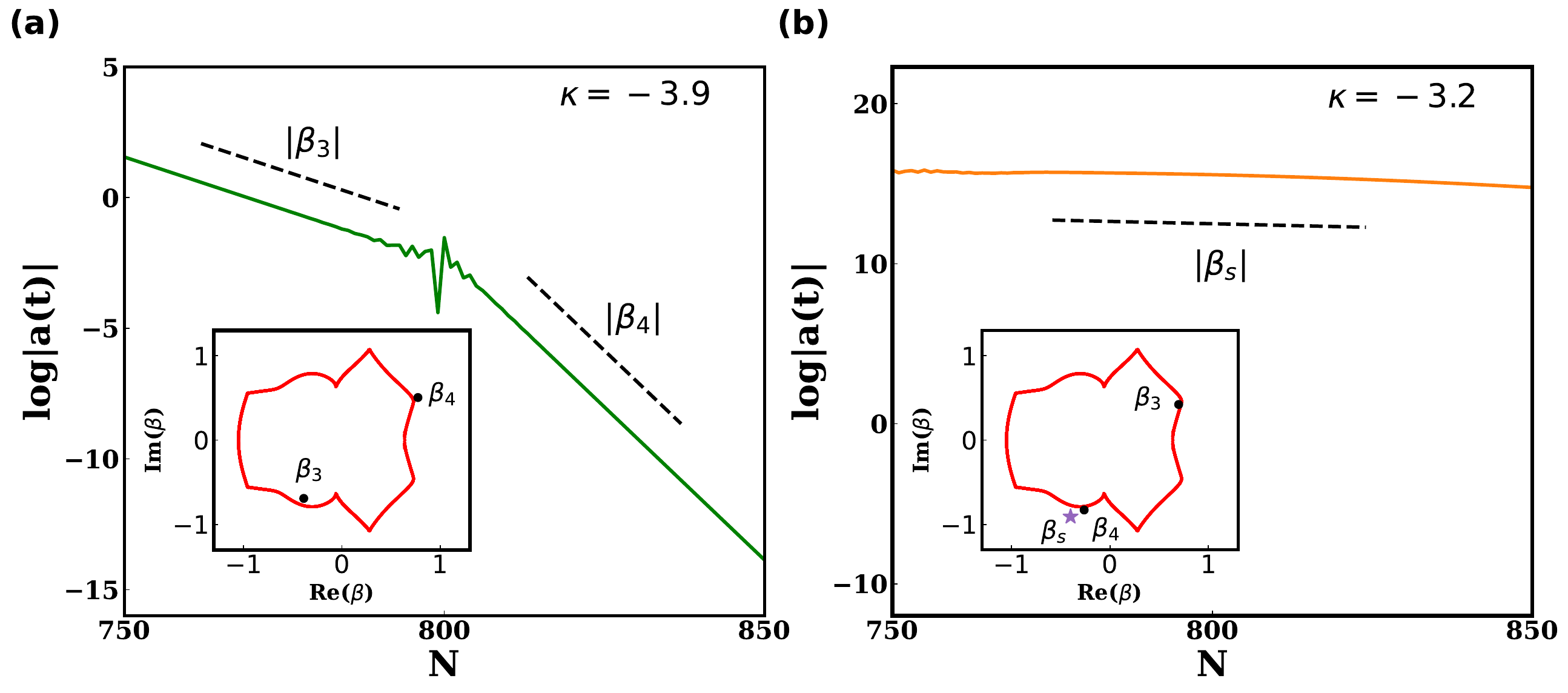}
    \caption{Spatial distribution of response near the excitation site 800 at $t=100$ for model Eq.~\ref{eq-model} with different on-site dissipation (a) $\kappa=-3.9$ and (b) $\kappa=-3.2$. Inset in (a) shows the two roots $\beta_3(\omega),\beta_4(\omega)$ of forced frequency $\omega$ close to GBZ. Besides these two roots, inset in (b) also shows $\beta_s$ at USP.}
    \label{fig-3}
\end{figure}

After deriving the response-excitation relationships, we can now explain the results in Fig.~\ref{fig-2} in detail. When $\kappa = -4.9 $ and $\kappa = -3.9 $, the saddle points' frequencies below $\operatorname{Im}\left(E_m\right)$ have only negative imaginary parts, 
regardless of whether $\operatorname{Im}\left(E_m\right)$ is above the real axis. Consequently, the natural response decays and the forced response survives. However, when $\kappa = -3.2$, the frequency of USP has a positive imaginary part, then the amplitude is exponentially growing characterized by this frequency. 

The forced response and the natural response also exhibit significant difference in spatial distribution as shown in Fig.~\ref{fig-3}. When $\kappa = -4.9$, the forced response dominates. The response at the right side of the excitation site $[G(t)]_{ij}(E) \sim\beta_4^{i-j}$, while at the left side $[G(t)]_{ij}(E) \sim \beta_3^{i-j}$, in agreement with the theoretical prediction by Eq.~\eqref{eq-GBZ}, demonstrating a potential application as directional amplifiers. However, when $\kappa = -3.2$, the natural response governs the dynamic evolution. The numerical calculation of spatial distribution of response on both sides of the excitation point follows the same scaling relation $[G(t)]_{ij}(I)\sim \beta_s^{i-j}$, consistent with Eq.~\eqref{eq-saddle}. These results also confirm the excitation–response relationships from another perspective.

\begin{figure}
    \centering
    \includegraphics[width=8.6cm]{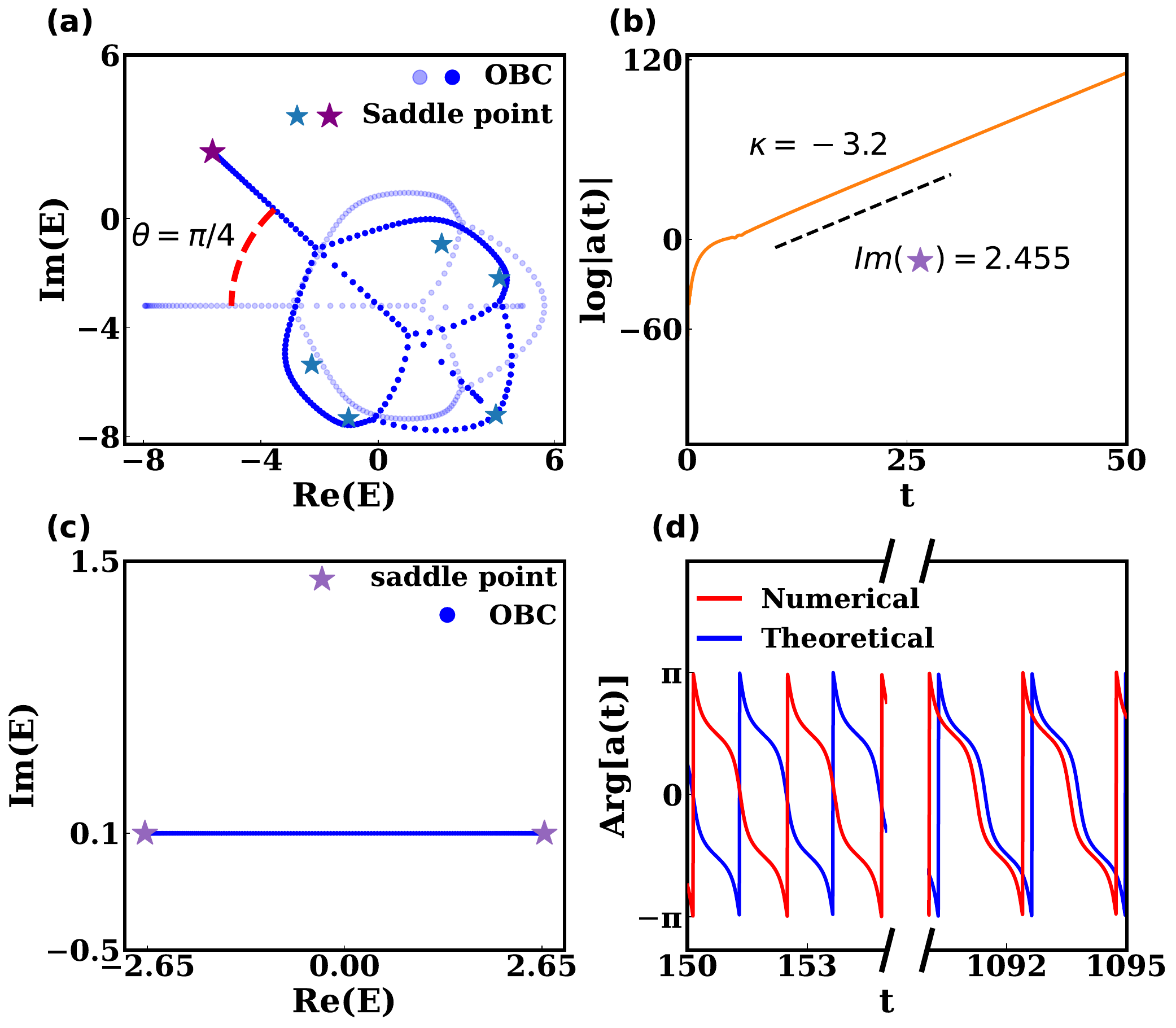}
    \caption{(a) OBC spectrum of corrected model Eq.~\ref{eq-model} with each parameter multiplied by $e^{-i\frac{\pi}{4}}$. (b) The dynamic evolution of amplitude aligns with the frequency of new USP in (a). (c) The OBC spectrum and (d) the variation of augment of amplitude with different time intervals for corrected Hatano-Nelson model $h(\beta)=1.8\beta+\beta^{-1}+0.1i$. The numerical result is denoted by red line and the theoretical result is denoted by blue line. Here, the excitation and response site are located at positions 2000 and 1950, respectively, in a lattice of length $N=4000$.}
    \label{fig-4}
\end{figure}

\textit{The competition among saddle points.}--- The general formulas of excitation–response relationships also explain the observation 
of $ \operatorname{Im}\left(E_m\right)$-dominated dynamic evolution of probabilities of the photon experimentally since this is where $ \operatorname{Im}\left(E_m\right)$ corresponds to the frequency of USP~\cite{PhysRevLett.126.230402,PhysRevLett.132.050402}. Such a scenario has its counterpart in corrected model Eq.~\eqref{eq-model} by multiplying the same factor $e^{-i\frac{\pi}{4}}$ on each parameter, as shown in Fig.~\ref{fig-4}(a). Here, the OBC spectrum is rotated by 45 degree and the imaginary part of USP is $\operatorname{Im}\left(E_m\right)$. Considering the same excitation and response site as in Fig.~\ref{fig-2}, the response amplitude is  exponential growth with frequency matching that of this new USP as shown in Fig.~\ref{fig-4}(b). This result reflects the competition among saddle points as a source of response frequency. 

We further investigate the scenario where the OBC spectrum possesses the same positive imaginary parts as exemplified by the corrected Hatano-Nelson model in Fig.~\ref{fig-4}(c). Here, the two ends of OBC spectrum are the frequencies of saddle points and both saddle points are USP. Unlike the previous focus on the temporal growth of the amplitude, here we investigate the argument of response amplitude, as shown in Fig.~\ref{fig-4}(d). The theoretical results involving the superposition of contribution from these two saddle points given by Eq.~\eqref{eq-saddle} are essentially consistent with the numerical results in the oscillation period. Figure~\ref{fig-4}(d) also shows the phase difference between theoretical and numerical results. This difference decreases as time increases since it takes a considerable period for the asymptotic solutions Eq.~\eqref{eq-saddle} to reach the numerical results, as shown in Fig.~\ref{fig-4}(d). The reason for this can be found in the Supplemental Material~\cite{SupplementalMaterials}.

\textit{Conclusions and discussions.}---Utilizing the Laplace transform in the dynamic equation, we theoretically derive general formulas of excitation–response relationships for non-Hermitian crystal systems under periodic excitation. Our results indicate the response contains the forced response from external driving and the natural response from frequencies of system. Intriguingly, the latter initially consists of OBC spectrum but evolves towards the saddle point below the  $ \operatorname{Im}\left(E_m\right)$ at long times. Such a transition suggests that for modes with frequencies containing positive imaginary parts, their excitations may dissipate during time evolution rather than leading to the  exponential growth in population as commonly understood. Consequently, the long-time behavior of response is determined by the competition between forced response and natural response from these saddle points. 

The frequencies of these saddle points can lie within or beyond the OBC spectrum, i.e., the corresponding wavevectors are always located on the aGBZ and may not be found on the GBZ, suggesting additional applications of aGBZ beyond its initial purpose of calculation of the GBZ. When the frequencies of all these saddle points lie below the real axis, the forced response dominates; otherwise, the system's amplitude experiences exponential growth with the frequency of USP, leading to instability. Thus, our theory provides a comprehensive picture of the dynamic evolution and offers insights into stability criterion of non-Hermitian crystal systems.

Given that non-Hermitian physics introduces many unique properties distinct from its Hermitian counterpart, we envision that it will significantly advance both the linear and nonlinear response theory. For example, if single higher-order saddle point with positive imaginary part exists on the OBC spectrum, the evolution process dominated by the OBC spectrum to saddle point frequency could be totally different due to the unique time-dependent response characteristics~\cite{ablowitz2003complex}. Another example involves the complexity of multivalued energy and the intricate substructure of multiband GBZs in multiband systems, where the forced response is attributed to the sum of contributions from different subbands~\cite{PhysRevResearch.5.043073}. Additionally, multi-band GBZs are typically highly dependent on specified perturbations, suggesting a unique response to perturbation. Another interesting direction for further investigation is the impact of exceptional points on response and related applications due to the  coalescence of eigenvector 
and their sensitivity to the perturbations~\cite{RevModPhys.93.015005}, which has recently led to the observation of a super-Lorentzian frequency response in non-Hermitian resonant systems with exceptional points~\cite{hashemi2022linear}.

This work is supported by the
Natural Science Foundation of Hunan Province ((Grant No. 2024JJ6011) and Innovation Program for Quantum
Science and Technology (Grant No. 2021ZD0302300).

\bibliography{manuscript.bib}

\end{document}